\documentclass[10pt,twocolumn,letterpaper]{article}
\usepackage[T1]{fontenc}
\usepackage[utf8]{inputenc}
\usepackage[margin=0.75in,columnsep=0.25in]{geometry}
\usepackage{newtxtext,newtxmath,microtype}
\usepackage{graphicx,booktabs,tabularx,array}
\usepackage{amsmath}
\usepackage[numbers,sort&compress]{natbib}
\usepackage{algorithm,algpseudocode}
\usepackage{xcolor,url,placeins,balance}
\usepackage[font=small,labelfont=bf]{caption}
\usepackage{fancyhdr}
\usepackage[colorlinks=true,linkcolor=teal,citecolor=teal,urlcolor=teal,breaklinks=true]{hyperref}
\hypersetup{pdftitle={AMP Atlas: A Browser-Only Workbench for Antimicrobial Peptide Exploration and Interactive Helical-Wheel Design},pdfauthor={Truong-Son Hy}}
\graphicspath{{figures/}}
\newcommand{\AAset}{\mathcal{A}}
\newcommand{\ind}{\mathbf{1}}
\newcommand{\code}[1]{\texttt{#1}}
\newcommand{\BenchmarkRows}{%
Partial wheel & 7 & 0.84 & 0.84--1.14 \\
Filtered wheel & 7 & 0.81 & 0.80--0.83 \\
Sequence neighbors & 3 & 878.66 & 875.44--886.72 \\
Descriptor neighbors & 3 & 20.08 & 19.38--20.45 \\
Structure neighbors & 3 & 60.43 & 60.33--64.39 \\
}

\begin{document}
\twocolumn[{
\begin{center}
{\LARGE\bfseries AMP Atlas: A Browser-Only Workbench for\\[3pt]
Antimicrobial Peptide Exploration and\\[3pt]Interactive Helical-Wheel Design\par}
\vspace{12pt}
{\large Truong-Son Hy}\par\vspace{5pt}
Department of Computer Science, University of Alabama at Birmingham, United States\\[3pt]
\href{mailto:thy@uab.edu}{thy@uab.edu}\\[7pt]
18 October 2026
\end{center}
\begin{abstract}
Accessible research software can connect published antimicrobial peptide (AMP) records to sequence analysis, molecular visualization, and interactive design without requiring a dedicated application server. We present AMP Atlas, an open-source, browser-only workbench that combines a DRAMP-derived database with local physicochemical calculations, published-neighbor retrieval, and sequence-guided structural comparison. The reported snapshot contains 12,784 source records, 10,579 canonical records, 5,183 source SMILES annotations, and 433 PDB chain structures. An inverse helical-wheel editor lets users specify peptide length, fill residues gradually, and immediately retrieve exact-length database matches under positional and property constraints. Selected-position residue frequencies support further editing while preserving the remaining constraints. A Web Worker performs searches and numerical calculations; a WebGL viewer provides interactive molecular representations. The static distribution occupies 138.8 MB and supports GitHub Pages hosting. Verification includes 23 Python tests, 30 browser-core/worker checks, a designer DOM integration suite, and agreement against fixed numerical references. This report describes the implemented algorithms and assumptions; it does not establish predictive accuracy for antimicrobial activity or experimentally validate designed peptides. The website is available at \url{https://hysonlab.github.io/amp-atlas/}, and the source code and database are available at \url{https://github.com/HySonLab/amp-atlas}.
\end{abstract}
\vspace{8pt}
}]

\section{Introduction}
Research on antimicrobial peptides brings together sequence records, chemical modifications, physicochemical descriptors, literature annotations, and experimentally determined structures. For an individual researcher, moving between these representations can require several tools and repeated transfers of the same sequence. A lightweight, inspectable workbench can make published results easier to explore, compare, and communicate.

DRAMP 5.0 motivates this integration by enriching AMP records with chemical information and introducing FASTS, which places a SMILES representation in a FASTA-style annotation line \citep{dramp2026}. AMP Atlas uses a downloaded general-record collection and associated source SMILES annotations; it does not reproduce every DRAMP collection or replace its curation. The distinction matters because a canonical sequence alone may omit chemical features that affect a peptide's properties.

Helical-wheel projections provide another bridge between sequence and design. The classical representation places successive residues around an ideal helical axis \citep{schiffer1967}. Hydrophobic moments quantify the asymmetry of the resulting hydrophobicity distribution \citep{eisenberg1982}, and tools such as HELIQUEST support analysis of sequences with specified helical properties \citep{gautier2008}. AMP Atlas adds an inverse interaction: users fill numbered wheel positions, inspect the corresponding linear sequence, and retrieve published records compatible with a partially specified design.

The system is intended for exploratory research and preparation of hypotheses. Its contribution is the integration of a portable data snapshot, transparent calculations, molecular views, and incremental constraint-based search in a static deployment. This report documents that implementation \citep{atlas2026}, rather than proposing a new biochemical descriptor, alignment algorithm, or learned AMP predictor. In particular, the current software contains no trained model for activity, toxicity, minimum inhibitory concentration (MIC), helicity, or three-dimensional structure prediction.

\section{System overview and scope}
The public edition is a static website. HTML, CSS, JavaScript, JSON, PDB coordinates, and downloadable data files are served by GitHub Pages. Scientific calculations and searches execute on the visitor's device. There is no application server, account system, remote inference endpoint, or API key required for visitors. An optional Python implementation remains available as a local reference and local API, but it is not required by the hosted edition.

\begin{figure*}[t]
\centering\includegraphics[width=\textwidth]{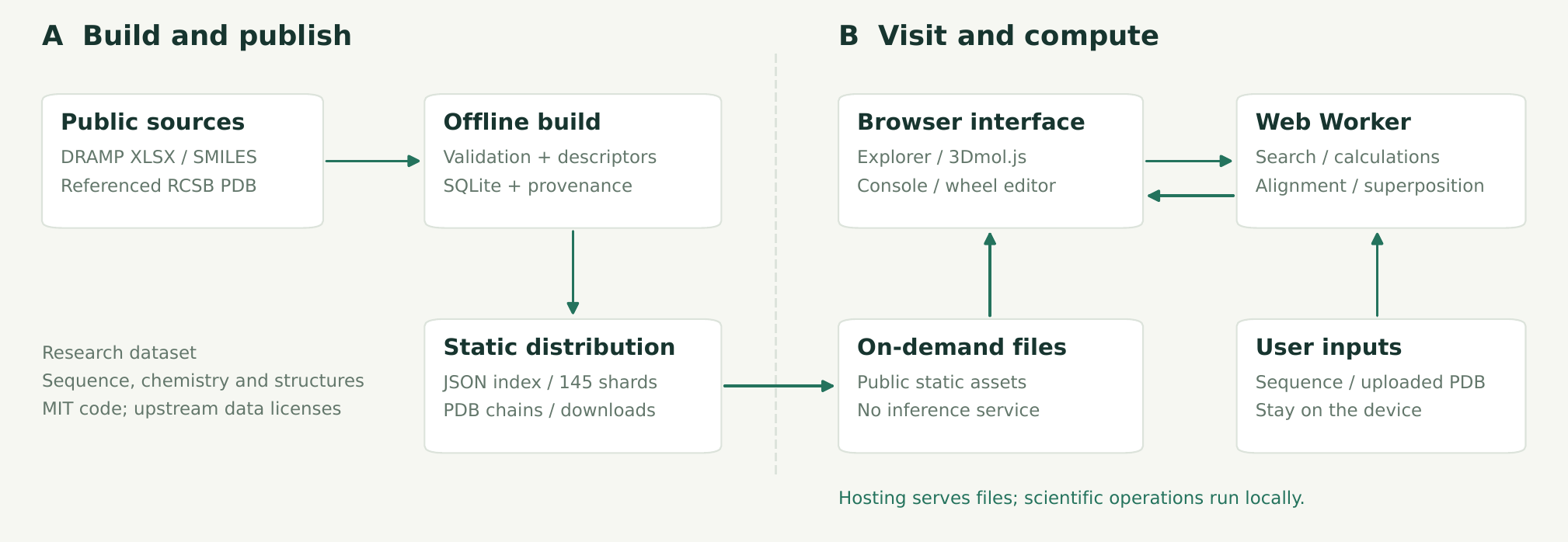}
\caption{System architecture. Data preparation occurs offline; hosting distributes a fixed set of public files. The main browser thread manages the interface and molecular viewer, while a Web Worker owns searchable data and executes numerical operations. User-provided sequences and PDB contents are processed locally. Arrows indicate data flow, not measured network traffic.}
\label{fig:architecture}
\end{figure*}

Figure~\ref{fig:architecture} separates the offline data preparation workflow from visitor-side operation. The interface exposes five connected workflows: browsing source records, analyzing an input sequence, comparing structures, editing a helical wheel, and downloading data. A completed wheel design can enter the same analysis workbench used for typed sequences; published results can return to the explorer. This reuse keeps literature evidence and design inputs connected.

Local processing avoids transmitting input sequences and uploaded PDB contents through the application's scientific operations. This is a statement about the implemented data flow, not a formal privacy audit: hosting providers can log ordinary page and public-file requests, and following an external publication link contacts that provider. The current site requires HTTP(S) for workers and file fetches; it is not a guaranteed fully offline application.

\section{Data collection and representation}
\subsection{Data sources and collection}
The database was assembled from official DRAMP general-record and general-SMILES spreadsheets downloaded on 17 October 2026, together with referenced PDB files obtained through RCSB PDB \citep{dramp2026,berman2000}. Source provenance and collection records are retained in the supplementary materials.

Table~\ref{tab:data} summarizes the dataset used in this study. Counts refer to imported records or stored chains, as specified, and should not be interpreted as counts of unique chemical entities. For example, 11,718 distinct source sequence strings include noncanonical strings, whereas only 10,056 distinct sequences occur among canonical records. A PubMed identifier count is obtained from source annotation text; it is not an independently verified bibliography.

\begin{table}[t]
\caption{Dataset composition, evaluated on 18 October 2026. Sizes use decimal MB ($10^6$ bytes).}
\label{tab:data}\centering\small
\begin{tabularx}{\linewidth}{@{}Xr@{}}\toprule
Quantity & Value\\\midrule
DRAMP source records & 12,784\\
Distinct source sequence strings & 11,718\\
Canonical records, length 1--500 & 10,579\\
Distinct canonical sequences & 10,056\\
Canonical records passing study heuristic & 10,489\\
Records with source SMILES & 5,183\\
Distinct source PubMed identifier strings & 3,452\\
Stored PDB chain structures & 433\\
Distinct PDB entries represented & 378\\
Records flagged with a linked structure & 315\\
Annotation shards & 145\\
Search index / structure index & 9.33 / 0.95 MB\\
SQLite database / complete static site & 42.55 / 138.81 MB\\\bottomrule
\end{tabularx}
\end{table}

\subsection{Validation and chemical annotations}
Source sequences are retained as provided. A separate canonical flag requires a sequence of 1--500 residues over the 20-letter alphabet
\[
\begin{split}
\AAset=\{&\mathrm{A,C,D,E,F,G,H,I,K,L,}\\
&\mathrm{M,N,P,Q,R,S,T,V,W,Y}\}.
\end{split}
\]
Noncanonical records remain browsable and downloadable; they are not silently converted to canonical peptides for calculation. Console input is uppercased and stripped of whitespace, then validated. FASTA headers and nonstandard residue symbols are rejected by the sequence console.

C-terminal annotations are mapped to amidated, free, or unknown status. Precomputed descriptors use the annotated amidation state when recognized; unknown status is calculated with free termini as an explicit approximation. Metadata retains source modification text and source SMILES. The software does not reconstruct covalent chemistry from SMILES, validate stereochemistry, or derive chemical identity from sequence equality. One imported row lacks a DRAMP identifier and is retained under the source-derived string \code{None} in an unassigned shard, rather than inventing an accession.

\subsection{Structure acquisition and linking}
The build report records 591 attempted referenced PDB downloads, including two failed downloads. The collection is therefore a bounded, source-referenced snapshot, not a complete structure survey. The parser reads the first PDB model, retains canonical \code{ATOM} C$\alpha$ positions, accepts blank or A alternate-location identifiers, and requires finite coordinates. Canonical chains with 5--150 observed residues are eligible for inclusion.

A candidate chain must match a referenced peptide's canonical sequence exactly or match a contiguous observed fragment covering at least 80\% of that peptide. A representative source record is selected for each retained chain, and relevant first-model atom records are saved for visualization. Display associations additionally use exact observed sequence equality across records. These rules produce 433 stored chains and 315 source records flagged with a local structure. They do not establish matching terminal chemistry, modifications, stereochemistry, or assay conditions. Missing coverage can reflect parser and collection rules as well as absent experimental structures.

\subsection{Storage and distribution}
SQLite stores peptide records, calculated properties, structure records, and provenance, with indexes on names, sequence, length, and structure associations. The static build emits a searchable JSON summary index, a structure index, and 145 annotation shards. Full annotations are loaded on demand, with at most 12 shard promises retained in the worker's insertion-ordered cache. PDB atom files are fetched when required for viewing. SQLite and bulk exports are downloaded only when requested.

The complete static distribution occupies approximately 138.8 MB, including downloadable exports. The approximately 9.33 MB uncompressed search index and 0.95 MB structure index load separately from detailed records and atom files, reducing the amount of data required for initial exploration. Transfer compression depends on the host.

\section{Interface and research workflows}
\subsection{Record exploration and molecular visualization}
The explorer searches record identifiers, names, sequences, and source-organism text, with pagination and filters for short peptides and linked structures. Selecting a record shows its sequence, calculated descriptors, original activity-related annotations, references, modification text, and available source SMILES. Links preserve a route back to DRAMP, PubMed, and structure providers.

The viewer embeds 3Dmol.js, a WebGL molecular visualization library \citep{rego2015}. Controls expose cartoon, sticks, spheres, and line representations, optional surfaces, residue/atom identification, labels, rotation, and image or coordinate export. The rendered molecular geometry comes from stored PDB atoms, not from the ideal wheel. Associated structures are labeled as sequence-matched references to avoid implying experimental determination of a user's new design.

\subsection{Sequence workbench}
Users enter a canonical sequence, choose pH and terminal amidation, and request either sequence-based or descriptor-based neighbors. Output includes descriptor cards, amino-acid composition, a residue hydropathy profile, a charge--pH curve, an ideal-helical projection, alignments, and source-study links. Up to 15 distinct-sequence neighbors are returned. These are nearest records according to the selected numerical rule, not predicted successful analogues.

\begin{figure*}[t]
\centering\includegraphics[width=\textwidth]{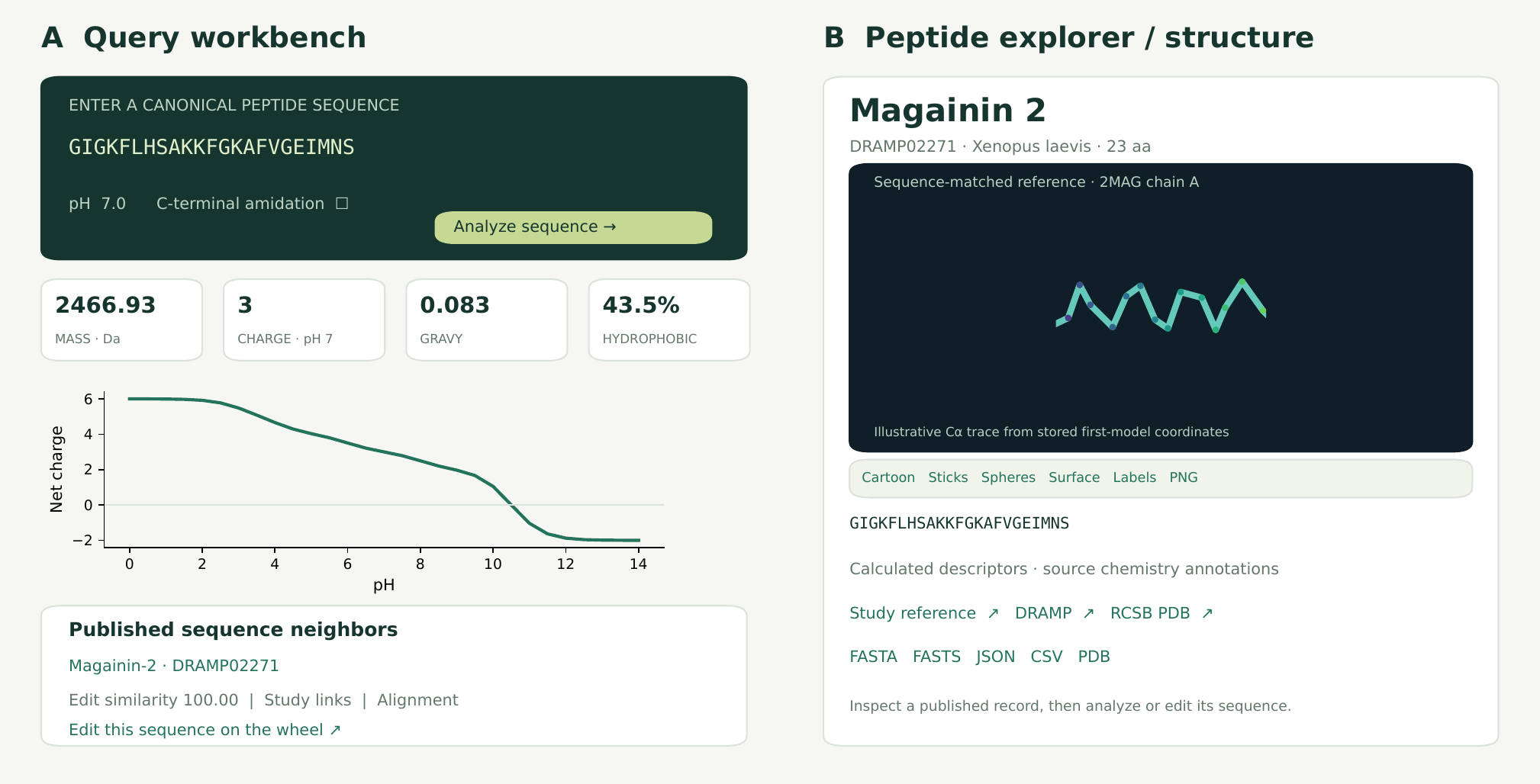}
\caption{Source-derived interface illustrations for the sequence workbench (A) and peptide explorer (B), using the actual magainin 2 sequence, descriptor outputs, and stored PDB 2MAG chain A data \citep{gesell1997}. These are vector reconstructions of selected interface components, not browser screenshots. The molecular panel shows an illustrative projection of stored C$\alpha$ coordinates; the live site uses the all-atom 3Dmol.js viewer \citep{rego2015}. The illustrated sequence is 23 canonical residues; this does not imply that every deposited residue or chemical group is represented in the descriptor model.}
\label{fig:workbench}
\end{figure*}

\subsection{Inverse helical-wheel designer}
The designer accepts lengths from 1 to 72 residues. Each numbered slot contains one canonical residue or an empty value. Users can click slots, choose from a residue palette, type residue letters, navigate with arrow keys, or clear a position with Delete. Filling advances in sequence order. Imports, clears, fills, and resizing participate in an undo history of up to 40 edits. Resizing preserves the shared sequence prefix.

The linear output reads positions in numbered N-to-C order. It never sorts residues by their plotted polar angle. Empty positions appear as question marks in the displayed pattern; they are search wildcards, not actual residues. Copy, FASTA export, and transfer to full-sequence analysis require every slot to be filled. SVG export is available for incomplete wheels as well.

Live search is enabled by default. After a 250 ms pause in editing, the current length, specified residues, selected position, and property filters are sent to the worker. A count summary appears beside the editor, and up to 100 matching source records are displayed. Users may pause automatic search and submit manually. Filters support inclusive lower/upper bounds for charge, GRAVY, and hydrophobic percentage, plus a linked-structure requirement. Invalid bounds produce a validation message.

The selected-position panel lists observed residue counts among records compatible with all other positions and filters. Clicking a residue applies it to the design. Counts are explicitly identified as database frequencies: duplicates, source selection, and publication bias can influence them. They are neither model-generated recommendations nor probabilities of activity.

\begin{figure*}[t]
\centering\includegraphics[width=\textwidth]{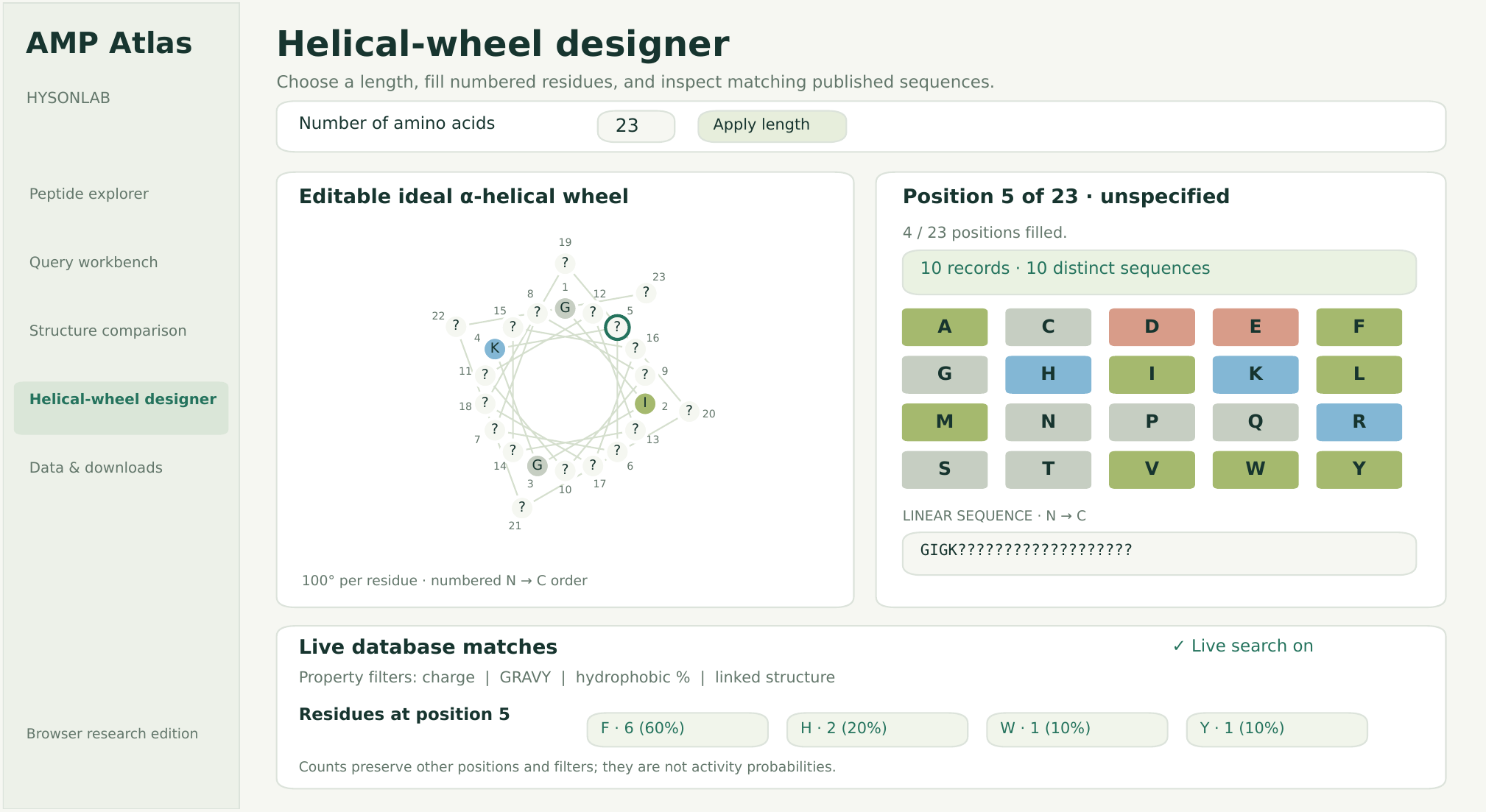}
\caption{Source-derived reconstruction of the live helical-wheel designer with length 23 and the first four residues set to GIGK. The actual worker returns 10 matching source records. Position 5 is selected; compatible residues are F (6), H (2), W (1), and Y (1). Geometry, colors, position numbering, pattern, and query outputs come from the current implementation. The report rearranges selected components for print; this is not a browser screenshot. Extra rings prevent slot overlap after residue 18 and do not represent a cyclic peptide.}
\label{fig:designer}
\end{figure*}

During partial entry, the interface reports how many assigned residues are basic (K/R/H), acidic (D/E), or hydrophobic (A/V/I/L/M/F/W/Y). It does not extrapolate a full mass, charge, or hydrophobic moment from a partially specified sequence. Once complete, the wheel panel displays mass, pH-7 charge, GRAVY, hydrophobic percentage, estimated pI, and ideal-helical moment under free-terminus assumptions. The workbench provides further analysis with its own pH and amidation controls.

\section{Algorithms}
\subsection{Sequence-derived descriptors}
Let $s=(a_1,\ldots,a_L)$ be a canonical sequence, $n_a$ its residue counts, and $I_{\rm am}$ indicate C-terminal amidation. The reported average molecular mass in daltons is
\begin{equation}
 M(s)=18.0153+\sum_{i=1}^{L}m(a_i)-0.9848I_{\rm am},
\end{equation}
where $m(a)$ is the residue mass listed in Appendix~\ref{app:parameters}. This canonical linear-chain model omits disulfide oxidation, cyclization, nonstandard residues, N-terminal modifications, and adducts.

Charge uses independent Henderson--Hasselbalch terms. With positive groups $\mathcal P=\{\mathrm{N\ terminus,K,R,H}\}$ and negative groups $\mathcal N=\{\mathrm{C\ terminus,D,E,C,Y}\}$,
\begin{equation}
 Q(p)=\sum_{g\in\mathcal P}\frac{n_g}{1+10^{p-pK_g}}
 -\sum_{g\in\mathcal N}\frac{n_g}{1+10^{pK_g-p}}.
\end{equation}
Terminal group counts are one, except the C-terminal contribution is omitted for amidation. The fixed p$K$ values are implementation parameters (Appendix~\ref{app:parameters}), not environment-dependent estimates. The charge curve samples pH 0--14 in 0.5-unit steps. Estimated pI uses 45 bisection iterations over $[0,14]$; when no zero is bracketed, the calculation approaches an interval boundary and should not be interpreted as a measured pI.

For Kyte--Doolittle hydropathy values $h(a)$ \citep{kyte1982}, mean hydropathy (GRAVY) is
\begin{equation}
 H(s)=\frac{1}{L}\sum_{i=1}^{L}h(a_i).
\end{equation}
The ideal-helical hydrophobic moment follows the vector-sum construction \citep{eisenberg1982}, using the same, unnormalized Kyte--Doolittle values:
\begin{equation}
 \mu_H(s)=\frac{1}{L}\sqrt{\left[\sum_i h(a_i)\cos\theta_i\right]^2+
 \left[\sum_i h(a_i)\sin\theta_i\right]^2},
\end{equation}
where $\theta_i=(i-1)100^\circ$. Its numerical scale depends on the chosen hydropathy table; it is not interchangeable with moments calculated using another scale or normalization. Neither this descriptor nor a wheel predicts helicity.

Hydrophobic fraction is $100\sum_{a\in\{\mathrm{A,V,I,L,M,F,W,Y}\}}n_a/L$; aromaticity uses F/W/Y. The reduced-cysteine extinction estimate is $5500n_{\rm W}+1490n_{\rm Y}$ in $\mathrm{M}^{-1}\mathrm{cm}^{-1}$, corresponding to the aromatic terms of the standard sequence-based coefficient model \citep{pace1995}. Cystine terms are not included. Composition and the unsmoothed per-residue hydropathy profile are also returned.

\subsection{Published-neighbor eligibility}
Similarity and wheel searches require a canonical record with a source-study annotation. The implementation accepts either a six-to-nine-digit token in the PubMed field, or a title longer than 12 characters together with a reference longer than six characters, excluding the placeholder titles ``not included yet'' and ``not found.'' This is a pragmatic annotation heuristic, not manual verification of each publication or its relevance. It admits 10,489 canonical records in the reported snapshot.

The explorer retains all source records. Neighbor lists deduplicate by exact sequence after ranking and return the first representative, with identifier ordering resolving score ties. Identical sequences can correspond to different modifications, organisms, or experiments; representative selection does not merge their chemistry or evidence.

\subsection{Global sequence similarity}
Sequence retrieval uses unit-cost Levenshtein distance. For sequences $s$ and $t$, the dynamic program is
\begin{align}
 D_{i,0}&=i,\qquad D_{0,j}=j,\\
 D_{i,j}&=\min\!\left\{\begin{array}{l}
 D_{i-1,j}+1,\\ D_{i,j-1}+1,\\ D_{i-1,j-1}+\ind[a_i\ne b_j].
 \end{array}\right.
\end{align}
The browser evaluates distances with rolling unsigned-16-bit rows: time is $O(|s||t|)$ per candidate and working memory is $O(|t|)$. Full alignments retain the traceback table and prefer a diagonal step on ties. This is a global minimum-edit alignment \citep{wagner1974}; it is not a substitution-matrix or affine-gap protein alignment. The optional Python distance implementation uses a bit-vector method \citep{myers1999} but computes the same unit-cost distance.

The displayed score is
\begin{equation}
 S_{\rm seq}(s,t)=100\left(1-\frac{D_{|s|,|t|}}{\max(|s|,|t|)}\right).
\end{equation}
Alignment identity is the proportion of identical residue pairs over all alignment columns, including gaps. The workbench's sequence coverage is aligned residue-pair count divided by query length. These quantities have different denominators and are reported separately.

\subsection{Physicochemical neighbor score}
Descriptor retrieval uses the five-vector
\[
 z(s)=(L,M,Q(7),H,\mu_H)
\]
and scale vector $\sigma=(20,2000,3,1,1)$. The score is
\begin{equation}
 S_{\rm prop}(s,t)=\frac{100}{1+\sqrt{\frac15\sum_{k=1}^{5}
 \left(\frac{z_k(s)-z_k(t)}{\sigma_k}\right)^2}}.
\end{equation}
These scales are fixed engineering choices, not learned weights or dataset standard deviations. Similarity is not calibrated to functional equivalence. Query descriptors for retrieval always use pH 7, even when the workbench displays charge at another pH. Stored candidate descriptors use their snapshot terminal assumptions. Wheel property filters likewise operate on stored values, including their existing numerical rounding; they do not refit chemical models during filtering.

\subsection{Wheel geometry and inverse mapping}
The editor retains slots in sequence order. For zero-based index $i$, its displayed angle and radius are
\begin{align}
 \phi_i&=(100i-90)^\circ,\\
 r_i&=105+58\lfloor i/18\rfloor,\\
 (x_i,y_i)&=(c+r_i\cos\phi_i,\ c+r_i\sin\phi_i),
\end{align}
where $c$ is half the SVG view-box size. Successive rings avoid collisions because the ideal projection repeats every 18 residues. Pixel radii and rings are layout parameters; they are not physical helix dimensions or peptide cyclization constraints. Linearization is simply the indexed concatenation $(a_1,\ldots,a_L)$ after every slot is filled, so the inverse mapping is deterministic.

\subsection{Incremental constraint-based retrieval}
Let $w\in(\AAset\cup\{\star\})^L$ be a partial wheel, with $\star$ an unspecified position. Let $\mathcal D_L$ contain eligible database records of length $L$, and let $F(r)$ indicate that record $r$ passes all active descriptor bounds and the optional structure flag. The matching set is
\begin{equation}
 \mathcal M(w,F)=\{r\in\mathcal D_L:F(r)\land
 \forall i,\ w_i=\star\ \lor\ s_i(r)=w_i\}.
\end{equation}
At least one residue must be specified. The result contains both source-record and distinct-sequence counts, as well as the positional-match count before property filters. Thus, a zero after filtering can be distinguished from a zero caused by the sequence pattern. Returning the first 100 records limits display size, not the underlying match count.

For selected position $j$, form $w^{(-j)}$ by replacing $w_j$ with $\star$. The residue count and displayed percentage are
\begin{align}
 c_j(a)&=\sum_{r\in\mathcal M(w^{(-j)},F)}\ind[s_j(r)=a],\\
 f_j(a)&=100c_j(a)/|\mathcal M(w^{(-j)},F)|.
\end{align}
When the denominator is zero, no frequencies are shown. Otherwise, nonzero residues are sorted by descending count, then alphabetically. The current implementation scans records and accumulates counts, taking $O(N_L L)$ time in the worst case. There is no combinatorial enumeration of all $20^L$ possible designs.

\begin{algorithm}[t]
\caption{Live wheel search with stale-response protection}
\label{alg:live}
\begin{algorithmic}[1]
\State Initialize generation $g\gets0$ and no pending timer
\Procedure{Edit}{$w,j,F$}
\State $g\gets g+1$; cancel pending timer
\State Clear results belonging to the previous state
\If{$w$ is empty or bounds are invalid}
\State Show an input/status message; \Return
\EndIf
\If{automatic search is paused}
\State Wait for manual submission; \Return
\EndIf
\State Set a 250 ms timer carrying $(w,j,F,g)$
\EndProcedure
\Procedure{Execute}{$w,j,F,v$}
\State Request worker matches and selected-slot counts
\State Await the result or error
\If{$v=g$}
\State Render the returned result or error
\EndIf
\EndProcedure
\end{algorithmic}
\end{algorithm}

A monotonically increasing generation counter prevents an older response or error from overwriting a newer state (Algorithm~\ref{alg:live}). Debouncing reduces redundant calls during rapid editing. It does not terminate an already running worker computation: an older request may still consume time, but its response is ignored. Manual submission executes immediately. DOM and controller tests check pausing, cancellation of timers, late responses, and workbench handoff.

\subsection{Sequence-guided structure comparison}
A query is either a stored chain or a locally uploaded PDB file, limited to 2 MB. For uploads, the parser selects a requested chain or the longest eligible canonical chain in the first model. Structural queries require 8--150 observed canonical residues. Each target is globally aligned to the query sequence; all aligned residue pairs, including substitutions, define correspondences. Targets require at least eight paired positions and coverage
\begin{equation}
 C=\frac{K}{\max(L_q,L_t)}\ge0.8,
\end{equation}
where $K$ is the number of aligned residue pairs. There is no additional identity threshold.

For paired C$\alpha$ coordinates $q_i,t_i$, rigid fitting minimizes
\begin{equation}
 \min_{R\in SO(3),\,u\in\mathbb R^3}
 \frac{1}{K}\sum_{i=1}^{K}\|q_i-(Rt_i+u)\|_2^2.
\end{equation}
The browser constructs Horn's symmetric $4\times4$ quaternion matrix and obtains the eigenvector of its largest algebraic eigenvalue using Jacobi rotations \citep{horn1987}. The Python reference uses an SVD-based Kabsch proper-rotation fit \citep{kabsch1976}. Both prohibit reflection and scaling. RMSD is the square root of the minimized objective and is reported in angstroms. A fitted transform is applied to the full observed target trace for overlay.

The selected stored query chain is excluded as an exact identifier; targets are deduplicated by PDB identifier and observed sequence. Results are ordered by increasing RMSD, then decreasing coverage, then identifier, with at most 15 displayed. C$\alpha$ radius of gyration is the root mean squared distance to the coordinate centroid; end-to-end distance uses the first and last observed C$\alpha$ atoms. These are unweighted trace descriptors. The procedure is sequence-guided rigid comparison, not sequence-independent fold search, flexible superposition, or a test of shared biological function.

\section{Reproducible examples and verification}
\subsection{Progressive search example}
The running example is the 23-residue sequence
\[
 \texttt{GIGKFLHSAKKFGKAFVGEIMNS},
\]
associated with magainin 2 in record \code{DRAMP02271}. PDB 2MAG provides a structural reference from an NMR study in membrane-mimicking conditions \citep{gesell1997}; AMP Atlas retains only its first-model observed canonical trace for numerical comparison. The NMR environment and retained-model choice limit interpretation of any overlay.

\begin{figure*}[t]
\centering\includegraphics[width=\textwidth]{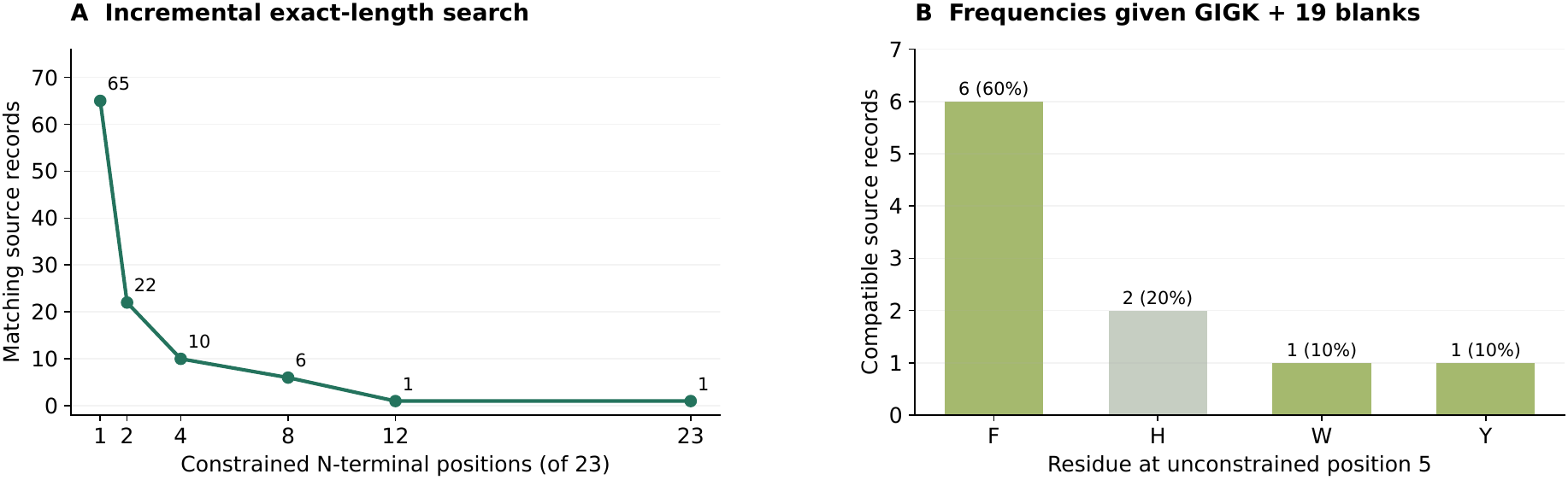}
\caption{Reproducible query results from the bundled data. (A) Match counts as an increasing N-terminal prefix of the 23-residue example is specified. (B) Position-5 residue counts with GIGK fixed and the remaining positions blank. Each source record contributes once, so percentages reflect this particular collection and its redundancy. No activity labels or predictive probabilities enter these calculations.}
\label{fig:search}
\end{figure*}

At fixed length 23 and with no property filters, constraining the first 1, 2, 4, 8, 12, and 23 positions yields 65, 22, 10, 6, 1, and 1 source records, respectively (Figure~\ref{fig:search}). Distinct-sequence counts are 64, 21, 10, 6, 1, and 1. After the prefix GIGK is entered, selected-position counts for residue 5 are F:6, H:2, W:1, and Y:1. These values are exact outputs of the deployed worker against the bundled snapshot, rather than simulated interface values. A separate pattern specifying only G at position 1 and K at position 4 yields two records when restricted to linked structures and nonnegative stored charge.

Under free-terminus, pH-7 assumptions, the example's calculated mass is 2466.93 Da, charge is 3.000, GRAVY is 0.083, hydrophobic fraction is 43.5\%, pI estimate is 10.501, and ideal-helical Kyte--Doolittle moment is 1.456. These outputs illustrate the implemented model, not measurements. The closest returned structural target for stored query \code{2MAG\_41} is \code{2FLY\_41}, with RMSD 1.1335 \AA\ and 87.0\% coverage under the stated correspondence rule. This nearest result is not evidence of identical chemistry or activity.

\subsection{Software and numerical verification}
The application has 23 Python tests and 30 browser-core/worker checks, supplemented by a designer DOM integration suite. Coverage includes validation, descriptors, global alignment, structure fitting, reflection rejection, static subdirectory paths, shard retrieval, pagination, exports, neighbor modes, PDB uploads, wheel geometry for all lengths 1--72, wildcard search, bounds validation, frequency counts, and stale-response handling. The DOM suite exercises actual designer controls and transfer into the existing analysis workbench using a DOM emulator. It does not validate real-browser rendering or WebGL behavior.

Across 30 fixed descriptor fixtures, the stored rounded Python and JavaScript descriptor outputs agree exactly for mass, charge, pI, GRAVY, hydrophobic fraction, moment, aromaticity, and reduced extinction coefficient. Across 20 fixed fitting fixtures, rounded RMSD outputs agree exactly and the maximum absolute fitted-coordinate discrepancy is $3.72\times10^{-14}$ \AA. The tests additionally reject a reflected configuration as a zero-error proper rotation. These checks support numerical consistency between implementations; they do not establish external biochemical accuracy.

Static integrity checks validate all 145 record shards, structure paths, database-copy equality, and asset paths. The application was also checked by the repository's verification workflow, and the corresponding published application assets were compared against the tested generated files. For this report, source-derived figures and algorithmic tests were used because automated browser capture was unavailable; no new visual cross-browser or WebGL compatibility study is claimed.

\subsection{Illustrative execution times}
Table~\ref{tab:timing} reports warm execution times for the unmodified deployed worker loaded into a Node.js VM against local files. The machine used Linux x64, Node.js v24.19.0, and an Intel Core i7-9750H CPU identified at 2.60 GHz. One untimed warm-up preceded each task's repeated measurements. Wheel timings use the length-23 pattern G at position 1 and K at position 4; the filtered case additionally requires a linked structure and nonnegative charge. Neighbor timings use the full example sequence; structure timing uses \code{2MAG\_41}.

\begin{table}[t]
\caption{Illustrative warm worker execution times in milliseconds. $n$ is the timed repetition count; ranges are minimum--maximum, not confidence intervals. These are Node.js measurements, not browser interaction or network latencies.}
\label{tab:timing}\centering\small
\begin{tabular}{@{}lrrr@{}}\toprule
Operation & $n$ & Median & Range\\\midrule
\BenchmarkRows
\bottomrule\end{tabular}
\end{table}

The measurements exclude initial JSON parsing/loading, DOM updates, molecular rendering, network transfers, and the deliberate 250 ms debounce. They use a small number of repetitions on one machine; no throughput, mobile-performance, or hardware-independent responsiveness claim follows. In particular, exhaustive global sequence comparison is substantially slower than wheel filtering at this scale, motivating future indexing work.

\section{Limitations and future work}
\paragraph{Chemical and biological scope.}
Canonical linear-chain descriptors cannot fully represent modified, cyclic, branched, or stereochemically altered peptides. Source SMILES improves evidence preservation, but it is not interpreted by the present calculator. Assay annotations remain source text; the software does not normalize MIC units, experimental organisms, media, toxicity protocols, or uncertainty. A matching record does not validate a designed peptide, and a missing match does not establish novelty or inactivity.

\paragraph{Structural interpretation.}
The corpus covers a small fraction of source records and is selected through available DRAMP references, download limits, and sequence-linking criteria. First-model observed traces omit ensemble variability and unobserved residues. Rigid sequence-guided RMSD is sensitive to correspondence, length, and coverage, and unrelated short helices can superpose well. No fold-recognition or binding-mechanism benchmark has been performed.

\paragraph{Evidence and evaluation.}
Annotation heuristics can admit incomplete or erroneous study references. Duplicate source records influence selected-position frequencies, while sequence-deduplicated neighbor lists can hide chemically distinct examples. Current evaluation establishes software consistency and reproducibility for a fixed snapshot. It includes no prospective peptide synthesis, biological assay, user study, retrieval-relevance judgment, or comparison of discovery outcomes against another platform. The report's interface figures are explicitly labeled source-derived reconstructions.

\paragraph{Scale and deployment.}
The static architecture reduces operational requirements for this collection, but the search index is loaded into each visitor's worker memory. Sequence retrieval scans all eligible records, and full alignments require quadratic tables for selected candidates. For larger collections, length/position inverted indexes, compact binary data, additional sharding, or optional WebAssembly kernels could improve execution and transfer cost. A future larger deployment should measure browser memory, actual interaction latency, cache behavior, and device variability. The current 138.8 MB deployment alone does not establish scalability to much larger data.

\paragraph{AI extensions.}
A future learned model would require explicitly licensed training data, normalization of labels and chemistry, sequence-aware train/test separation, calibrated uncertainty, and independent evaluation. Such models should remain separate from deterministic descriptors and source frequencies so that users can distinguish calculations, observations, and predictions. These are development directions, not implemented results.

\section{Availability and reproducibility}
AMP Atlas is freely accessible through its public website, with source code and data distributed through GitHub \citep{atlas2026}; both links are provided in the abstract. The application can also be run locally using a static HTTP server. SQLite, JSON, CSV, FASTA, and FASTS downloads support downstream analysis and reuse. FASTS exports retain source SMILES and source FASTS sequence text when present; this does not independently validate the encoded chemistry.

Supplementary materials provide the data provenance, exact query outputs, numerical evaluation results, and scripts for reproducing the reported analyses and figures. The repository includes the application build tools and verification suites. The evaluation harness executes the same scientific worker used by the website against locally stored data.

Application source is MIT licensed. DRAMP-derived material is attributed to its curators and distributed subject to its upstream CC BY 4.0 terms \citep{dramp2026}; PDB archive data follow CC0, with citations to structure depositions and associated studies \citep{berman2000,gesell1997}. Bundled visualization code retains its own license and notices. These upstream licenses remain distinct from the application license. AMP Atlas is an independent research workbench and is not an official DRAMP release.

\section{Conclusion}
AMP Atlas connects source peptide records, transparent sequence calculations, molecular inspection, and inverse helical-wheel editing in a browser-only application. The live designer treats partial sequences as explicit constraints and exposes the source records and selected-position counts behind its feedback. The available data, transparent algorithms, and numerical verification support reproducible exploration. The current contribution is a portable exploratory research tool; experimental assessment, chemical generalization, and predictive modeling remain future work.

\FloatBarrier
\bibliographystyle{unsrtnat}
\bibliography{references}

\balance
\appendix
\section{Calculation parameters}\label{app:parameters}
Table~\ref{tab:residues} lists the complete canonical residue parameter tables used by the implementation. Mass values are average residue masses, not free amino-acid or monoisotopic masses. The hydropathy column reproduces the Kyte--Doolittle scale \citep{kyte1982}. Other assumptions and units are defined in Section 5.

\begin{table}[ht]
\caption{Canonical residue parameters. $m$ is average residue mass in Da; $h$ is the Kyte--Doolittle hydropathy value.}
\label{tab:residues}\centering\small
\begin{tabular}{@{}crr@{\quad}crr@{}}\toprule
Residue & $m$ & $h$ & Residue & $m$ & $h$\\\midrule
A & 71.0788 & 1.8 & M & 131.1926 & 1.9\\
C & 103.1388 & 2.5 & N & 114.1038 & $-3.5$\\
D & 115.0886 & $-3.5$ & P & 97.1167 & $-1.6$\\
E & 129.1155 & $-3.5$ & Q & 128.1307 & $-3.5$\\
F & 147.1766 & 2.8 & R & 156.1875 & $-4.5$\\
G & 57.0519 & $-0.4$ & S & 87.0782 & $-0.8$\\
H & 137.1411 & $-3.2$ & T & 101.1051 & $-0.7$\\
I & 113.1594 & 4.5 & V & 99.1326 & 4.2\\
K & 128.1741 & $-3.9$ & W & 186.2132 & $-0.9$\\
L & 113.1594 & 3.8 & Y & 163.1760 & $-1.3$\\\bottomrule
\end{tabular}
\end{table}

Charge calculations use p$K$ values 8.0 (N terminus), 3.1 (C terminus), 10.5 (K), 12.5 (R), 6.0 (H), 3.9 (D), 4.1 (E), 8.3 (C), and 10.1 (Y). These are the fixed values in the reported implementation; no experimental fit was performed for this report. Candidate and query comparison assumptions should be considered when interpreting relative rankings.

\section{Figure provenance}
The report's figures are generated from the current interface geometry and actual worker output. The original wheel SVGs exported from the designer DOM are included as supplementary figure sources. The example molecular trace is derived from local PDB 2MAG chain A coordinates, with the original study cited in the caption. No source-paper figure is reproduced.

\end{document}